\documentclass[reprint,aip,showpacs,twocolumn,letterpaper,superscriptaddress]{revtex4}
\usepackage{amssymb,amsfonts,amsmath}
\usepackage{times}

\usepackage{graphicx}
\usepackage{tabularx}
\usepackage{bm}

\usepackage{amssymb,amsfonts,amsmath}

\newcommand{\hdot}{$\text{H}^-\cdot\cdot\cdot \text{H}_2$}
\usepackage{color}	

\usepackage{subfigure}
\usepackage{multirow}
\newcommand{\mgh}[1]{MgH$_{#1}$}

\newcommand{\amgh}{\mbox{$\alpha$-\mgh2}}

\newcommand{\zmgh}{\mbox{$\zeta$-\mgh2}}
\newcommand{\hydrogen}{H$_2$}
\newcommand{\hydride}{H$^-$}

\usepackage{amsmath}
\usepackage{amssymb}
\usepackage{graphicx}
\usepackage{subfigure}
\usepackage{multirow}

\begin{document}
\title{Metallization of Magnesium Polyhydrides Under Pressure}
\author{David C.\ Lonie}
\affiliation{Department of Chemistry, State University of New York at Buffalo, Buffalo, NY 14260-3000, USA}
\author{James Hooper}
\affiliation{Department of Chemistry, State University of New York at Buffalo, Buffalo, NY 14260-3000, USA}
\author{Bahadir Altintas}
\affiliation{Department of Chemistry, State University of New York at Buffalo, Buffalo, NY 14260-3000, USA}
\affiliation{Department of Computer Education and Instructional Technologies, Abant Izzet Baysal University, 14280, Golkoy, Bolu-Turkey}
\author{Eva Zurek}\email{ezurek@buffalo.edu}
\affiliation{Department of Chemistry, State University of New York at Buffalo, Buffalo, NY 14260-3000, USA}


\begin{abstract}
Evolutionary structure searches are used to predict stable phases with
unique stoichiometries in the hydrogen--rich region of the
magnesium/hydrogen phase diagram under pressure. MgH$_4$, MgH$_{12}$
and MgH$_{16}$ are found to be thermodynamically stable with respect
to decomposition into MgH$_2$ and H$_2$ near 100~GPa, and all lie on
the convex hull by 200~GPa. MgH$_4$ contains two H$^-$ anions and
one H$_2$ molecule per Mg$^{2+}$ cation, whereas the hydrogenic sublattices
of MgH$_{12}$ and MgH$_{16}$ are composed solely of H$_2^{\delta -}$
molecules. The high--hydrogen content stoichiometries have a large
density of states at the Fermi level, and the $T_c$ of MgH$_{12}$ at
140~GPa is calculated to be nearly three times greater than that of
the classic hydride, MgH$_2$, at 180~GPa.
\end{abstract}


\maketitle

\section{Introduction}

In 1935, Wigner and Huntington predicted that hydrogen, which exists
in the paired molecular state at ambient pressures and temperatures,
would become an alkali--metal--like monoatomic solid when compressed
to pressures exceeding 25~GPa \cite{Wigner:1935}. This turned out to
be a bit of an underestimate. Structure searches based upon density
functional theory calculations have identified a number of molecular
($ P \le 400$~GPa) \cite{Pickard:2007a} and quasi--molecular or atomic
($P = 0.5 - 5$~TPa) \cite{LiuMa:2012a, Mcmahon:2011a} structures in
the cold phase diagram, and experiments in diamond anvil cells show
that the insulating phase III with paired hydrogens is stable over a
broad temperature range and up to at least 360~GPa
\cite{Zha:2012a}. Recent experimental work at room temperature above
220~GPa, which showed a pronounced softening of one of the Raman
active molecular vibrons upon compression \cite{Howie:2012a}, and potential conductivity
\cite{Eremets:2011a} has generated much excitement
\cite{Amato:2012a}. The newly discovered phase IV of hydrogen is
thought to be a mixed structure composed of layers of molecular units,
as well as weakly bonded graphene--like sheets \cite{LiuMa:2012b,
  Pickard:2012a}.  Superconductivity at pressures of 450~GPa and
temperatures up to 242~K has been predicted in the molecular phase
\cite{Cudazzo:2008a}, whereas the superconducting transition temperature, $T_c$, may approach 764~K for monoatomic
hydrogen near 2~TPa \cite{Mcmahon:2011b}.

What are the structural motifs that compressed hydrogen adopts when
doped with an electropositive element? Theoretical work has predicted
the presence of hydridic H$^-$ atoms, H$_2^{\delta -}$ molecules,
\hdot\ motifs, symmetric H$_3^-$ molecules, polymeric
(H$_3^-$)$_\infty$ chains, and sodalite cage structures, with the
nature of the hydrogenic sublattice depending upon the identity of the
alkali or alkaline earth metal, and the pressure \cite{Zurek:2009c,
  Zurek:2011d, Zurek:2011h, Zurek:2012b, Zurek:2012g, Zurek:2012n, Zhou:2012a,
  Ma:2012b}. The latter clathrate--like cage which encapsulated the
calcium cation in CaH$_6$ was shown to be susceptible to a Jahn--Teller
distortion, giving rise to a remarkable electron--phonon coupling
parameter of 2.69 at 150~GPa with a concomitant $T_c$ of $\sim$225~K
\cite{Ma:2012b}. 

Herein, evolutionary structure searches are used to seek out the
stoichiometries and structures of magnesium polyhydrides, MgH$_n$ with
$n>2$, under pressure. MgH$_4$ is predicted to become stable with
respect to decomposition into MgH$_2$ and H$_2$ near 100~GPa, and it
remains the most stable stoichiometry until at least
200~GPa. $Cmcm$--MgH$_4$ contains two hydridic hydrogens and one
hydrogen molecule per Mg$^{2+}$ cation, and it becomes metallic as a result
of pressure--induced band overlap. However, due to the low density of
states at the Fermi level, the estimated $T_c$ at 100~GPa is only
$\sim$10~K higher than that of $P6_3/mmc$--MgH$_2$ at 180~GPa. Two
other stoichiometries, MgH$_{12}$ and MgH$_{16}$, are calculated as
being thermodynamically stable at higher pressures. Because these
hydrogen rich phases exhibit an ``H$_2^{\delta -}$ belt'' surrounding
the Mg$^{2+}$ cations, and do not contain any hydridic hydrogens, they
have a high density of states at the Fermi level. Assuming typical
values of the Coulomb pseudopotential, the $T_c$ of MgH$_{12}$ is
calculated as being between 47-60~K at 140~GPa. Phases with a greater
ratio of H$_2^{\delta -}$:H$^-$ tend to have a larger density of
states at the Fermi level, and a higher $T_c$.

\section{Computational Methods}

The structural searches were performed using the open--source
evolutionary algorithm (EA) XtalOpt release 8 along with the default parameter
set \cite{Zurek:2011a}. Evolutionary runs were carried out on the
MgH$_2$ stoichiometry at 0, 25, 50, 100, 150, 200 and 250~GPa with
cells containing 2, 3, 4 and 8 formula units (FU).  At 200~GPa
additional searches on cells with 5 and 6 FU were performed.  Only the
known $\alpha$, $\epsilon$ and $\zeta$--phases were recovered, with
the $P6_3/mmc$ structure remaining the most stable up to the highest
pressures considered.  Exploratory searches at 200~GPa revealed that
MgH$_n$ with $n=2.5, 3, 3.5, 4.5, 5, ...$ were noticeably less stable
than those with even $n$. Thus, a more refined search was carried out
at 100 and 200~GPa which was restricted to even $n$ ranging from 4-16
with 2-4 FU for MgH$_4$ and MgH$_6$, 2-3 FU for MgH$_8$ and 2 FU
otherwise. Additional searches for $n=4$ and $n=6$ were carried out
with 8 and 5 FU at 100 and 200~GPa, respectively. Duplicate structures
were detected using the XtalComp \cite{Zurek:2011i} algorithm. The
spglib package \cite{web-spglib} was used to determine space--group
symmetries.

Geometry optimizations and electronic structure calculations were
performed using using density functional theory (DFT) as implemented
in the Vienna ab-initio simulation package (VASP)
\cite{Kresse:1993a}. The exchange and correlation effects were treated
using the Perdew-Burke-Ernzerhof (PBE) functional \cite{Perdew:1996a}
with plane--wave basis sets and a kinetic energy cutoff of 600 eV. The
hydrogen 1s$^1$ and magnesium 2p$^6$3s$^2$ electrons were treated
explicitly and the projector-augmented wave (PAW) method
\cite{Blochl:1994a} was used to treat the core states. The $k$--point
grids were chosen using the $\Gamma$--centered Monkhorst--Pack
scheme. The number of divisions along each reciprocal lattice vector
was chosen such that the product of this number with the real lattice
constant was 40~\AA{} for the final geometry optimizations, as well as
at least 50~\AA{} for the electronic densities of states (DOS) and band structures.

In situations where band gap closure occurs as a result of pressure induced broadening, and eventual overlap of the valence and conduction bands, standard density functionals predict too low metallization pressures. For this reason we have calculated the DOS of select structures using the HSE06 screened hybrid functional, which has been shown to give good accuracy for band gaps \cite{Krukau:2006a}. Due to the immense computational expense involved in hybrid calculations, the geometries employed have been optimized with PBE. Recently, it has been shown that inclusion of Hartree--Fock exchange in functionals such as HSE06 or PBE0 can have a significant impact upon the calculated transition pressures between different phases \cite{Liu:2012b, Teweldeberhan:2012a}. Moreover, a study considering liquid nitrogen at pressures up to 200~GPa (and finite temperatures) has shown that the structural relaxations with a hybrid functional can lead to large Peierls distortions (which inevitably have an impact on the band gap) \cite{Boates:2011a}. It has been proposed that nontrivial exchange correlation effects can become quite important at extreme pressures, in particular when electron localization occurs (``electride'' behavior \cite{Pickard:2009a}), and/or when the semicore electrons interact \cite{Teweldeberhan:2012a}. Since we do not observe significant broadening of the Mg $2p$ bands at the pressures considered here, it may be that structural relaxation with a hybrid functional will yield similar results as PBE. This will be considered in future studies. However, as expected, our calculations show that hybrid functionals increase the pressures at which band gap closure is predicted to occur.

Phonon calculations were performed using VASP combined with the
phonopy package \cite{phonopy} on supercells of 288 (MgH$_2$), 160 (MgH$_4$), 351
(MgH$_{12}$) and 272 (MgH$_{16}$) atoms. The electronic
densities of states and phonon band structures obtained with VASP for MgH$_2$, MgH$_4$ and MgH$_{12}$ at 180, 100 and 140~GPa showed good agreement to those computed using the Quantum Espresso (QE) program \cite{QE}, and the computational settings described below.

In the QE calculations the H and Mg pseudopotentials, obtained from the QE pseudopotential library, were generated by the method of von Barth and Car with 1$s^1$ and 3$s^2$3$p^0$ valence configurations, along with the Perdew--Zunger local density approximation \cite{PhysRevB.23.5048}. We choose this particular Mg pseudopotential as it has been employed in numerous lattice dynamical studies of materials under pressure, see for example Ref.\ \cite{Karki:2000a}. Plane wave basis set cutoff energies, which gave an energy convergence to better than 0.3~mRy/atom were 55, 75 and
90~Ry for MgH$_2$, MgH$_4$ and MgH$_{12}$, respectively, and a
$16\times16\times16$ Brillouin--zone sampling scheme of
Methfessel--Paxton \cite{PhysRevB.40.3616} with a smearing factor of
0.02~Ry was employed. Density functional perturbation theory, which is
implemented in QE, was used for the phonon calculations. The
electron--phonon coupling matrix elements were calculated using a
$16\times16\times16$ $k$--mesh and $4\times 4 \times 4$ $q$--mesh for
MgH$_2$ and MgH$_4$, along with a $12\times12\times12$ $k$--mesh and
$3\times 3 \times 3$ $q$--mesh for MgH$_{12}$. The electron phonon
coupling (EPC) parameter, $\lambda$, was calculated using a set of
Gaussian broadenings in steps of 0.005~Ry. The broadening for which
$\lambda$ was converged to within 0.005 was 0.035, 0.025
and 0.040~Ry for MgH$_2$, MgH$_4$ and MgH$_{12}$, respectively. The
superconducting transition temperature, $T_c$, has been estimated
using the Allen--Dynes modified McMillan equation
\cite{PhysRevB.12.905} as:
\begin{equation}
T_c = \frac{\omega_{\text{log}}}{1.2}
\text{exp}\left[-\frac{1.04(1+\lambda)}
  {\lambda-\mu^*(1+0.62\lambda)}\right]
\end{equation}
where $\omega_{\text{log}}$ is the logarithmic average frequency and
$\mu^*$ is the Coulomb pseudopotential, often assumed to be between
$\sim$0.1-0.13.

The molecular calculations on H$_2^{\delta -}$ and MgH$_{12}$ were
performed using the ADF software package \cite{adf} and the revPBE
gradient density functional. The band structures of select phases (see
the Supporting Information, SI) were calculated using the tight--binding linear muffin--tin
orbital (TB-LMTO) method \cite{Andersen:1984}, the VWN local exchange
correlation potential along with the Perdew--Wang GGA.

\section{Results and Discussion}

\subsection{Squeezing MgH$_2$}
\begin{figure}[b!]
\centering
\includegraphics[width=\columnwidth]{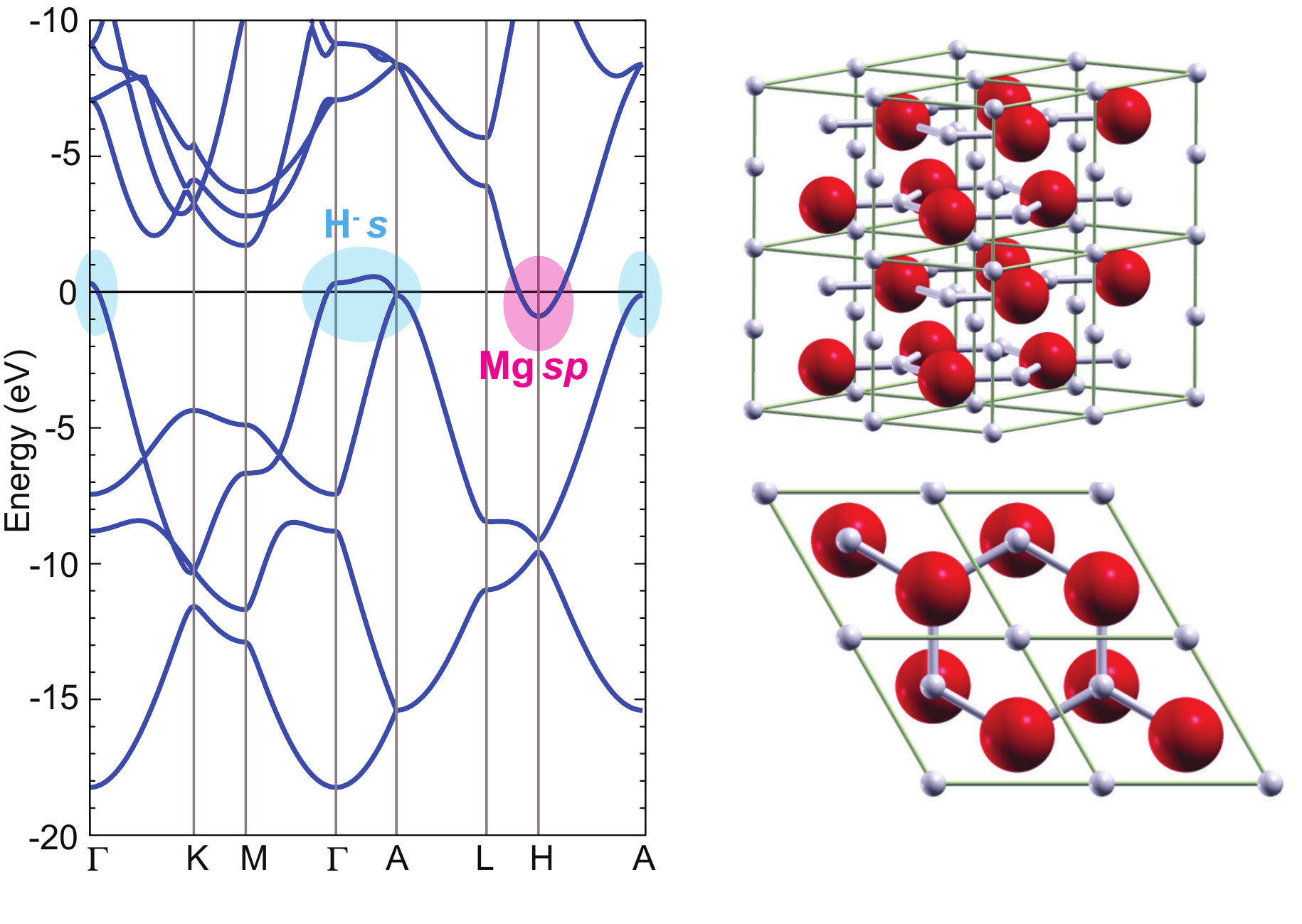}
\caption{The electronic band structure of \zmgh\ at 180~GPa. The
  character of the bands which cross the Fermi level is
  highlighted. The Fermi energy is set to zero in all of the plots. A
  side and top view of \zmgh\ is shown on the right, with the Mg/H atoms colored in red/white.}
\label{fig:MgH2}
\end{figure}

The potential for reversible hydrogen storage has resulted in much
interest not only in crystalline MgH$_2$, but also in nanoparticles
based on the structure of the various phases of this solid
\cite{Vajeeston:2012a, Vajeeston:2012b, Reich:2012a, Koukaras:2012a,
  Buckley:2012a}. At 1~atm \amgh\ assumes a TiO$_2$--rutile--type
geometry and it undergoes a series of structural transitions,
$\alpha~(P4_2/mnm) \rightarrow \gamma~(Pbcn) \rightarrow
\beta~(Pa\bar{3}) \rightarrow \delta~(Pbc2_1) \rightarrow
\epsilon~(Pnma)$, at 0.39, 3.84, 6.73 and 10.26~GPa
\cite{Vajeeston:2006a, Vajeeston:2002a}. At around 165~GPa the $Pnma$
phase slightly distorts to a higher symmetry $P6_3/mmc$ Ni$_2$In--type
configuration, \zmgh , see Fig.\ \ref{fig:MgH2}, which was shown to be dynamically stable at 180~GPa. At lower pressures
BaH$_2$ \cite{Tse:2009a, Chen:2010b}, CaH$_2$ \cite{Tse:2007b}, and
SrH$_2$ \cite{Smith:2009a} also adopt this structure. However, whereas
the heavier alkaline earth dihydrides with $P6_3/mmc$ symmetry are
insulating, in MgH$_2$ metallization is predicted to occur by 170~GPa
at the PBE--level of theory \cite{Zhang:2010b}. Our evolutionary runs
did not reveal any other phases of MgH$_2$ up to 300~GPa, a pressure
at which the metallicity in \zmgh\ persists. Calculations using the HSE06 screened hybrid functional
\cite{Krukau:2006a} confirm the metallicity at $\sim$180~GPa (see the SI).

\zmgh\ is metallic as a result of the closure of an indirect band
gap. A flat band displaying H$^-$ $s$--character rises above the Fermi
level, $E_F$, around $\Gamma - A$, and a steep band that boasts
primarily Mg $s$ and a little bit of Mg $p$--character falls below
$E_F$ around the $H$--point, as illustrated in
Fig.\ \ref{fig:MgH2}. \zmgh\ contains half the number of valence
electrons per formula unit as does MgB$_2$, which becomes superconducting at 39~K
\cite{Nagamatsu:2001a}. The B $\sigma$-- and H $s$--bands are
comparable, in particular the ``holes'' at the top of the band
spanning from $\Gamma$ to $A$. However, whereas MgB$_2$ consists of
hexagonal boron sheets with Mg$^{2+}$ intercalated in the hexagonal
holes, the hexagonal network in \zmgh\ is made up of alternating
Mg$^{2+}$ and H$^-$ ions, with the second set of hydrides located in
the hexagonal holes. So the Fermi surfaces arising from the B
$\sigma$-- \cite{Choi:2002a} and the H $s$--bands \cite{Zhang:2010b}
are not identical.  We also noticed some similarities between
\zmgh\ and the most stable CsH phase above 160~GPa
\cite{Zurek:2011e}. Whereas the former can be thought of as layers of
graphitic sheets of alternating Mg$^{2+}$ and H$^-$ ions arranged in
an ABABA... stacking with H$^-$ sandwiched in between the two layers,
in CsH half of the H$^-$ have been removed so that only layers of
Cs$^+$ and H$^-$ are found.

\begin{figure}
\centering
\includegraphics[width=\columnwidth]{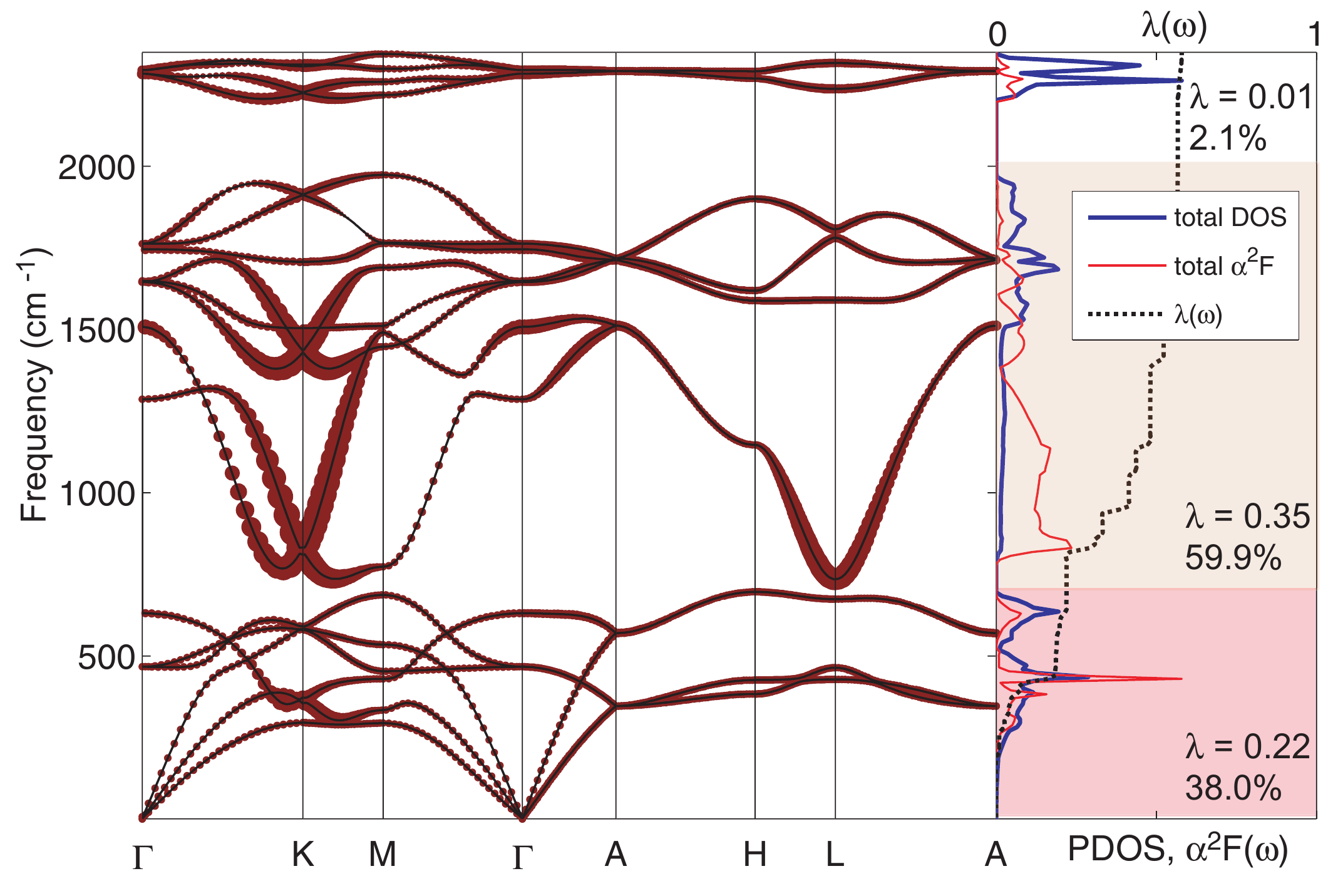}
\caption{Phonon band structure, phonon density of states and the
  Eliashberg spectral function, $\alpha^{2}$F($\omega$), of \zmgh\ at
  180~GPa. Circles indicate the phonon linewidth with a radius
  proportional to the strength.  At this pressure $\lambda=0.58$,
  $\omega_{log}=1111$~K, and $T_c=16-23$~K assuming
  $\mu^*=0.13-0.1$.} \label{fig:MgH2:critical}
\end{figure}

The density of states at the Fermi level, $g(E_F)$, of \zmgh\ is
0.01~eV~$^{-1}$/electron at 170~GPa and by 300~GPa it decreases only
slightly, see the SI. Despite the relatively
low $g(E_F)$ we were intrigued in the possibility of superconductivity
in this system. Fig.\ \ref{fig:MgH2:critical} plots the phonon band
structure and densities of states, phonon linewidths ($\gamma(\omega)$), Eliashberg spectral function ($\alpha^2
F(\omega)$), and $\lambda(\omega)$ of \zmgh\ at 180~GPa. 38\% of the
total EPC parameter, $\lambda$, is a result of the low--frequency modes
below $\sim$700~cm$^{-1}$ which are associated primarily with the motions of the heavier Mg ions, whereas the
region between 750-2400~cm$^{-1}$, which is mostly due to motions of the H$^-$ anions, contributes 62\% towards
the total $\lambda$ with the modes along $\Gamma-K-M$ and $H-L-A$
playing a dominant role. The modes above 2100~cm$^{-1}$ are primarily caused by the set of H$^-$ anions which are closer to the Mg$^{2+}$ cations (the Mg--H distances at 180~GPa measure 1.58 and 1.81~\AA{}).
Despite the modest $g(E_F)$
the total EPC parameter is calculated as being 0.58, and along with an
$\omega_\text{log}$ of 1111~K gives rise to a $T_c$ of 16-23~K for
$\mu^*$ ranging from $0.13-0.1$, respectively, via the Allen--Dynes
modified McMillan equation. For comparison, the EPC parameter and  $\omega_\text{log}$ for
BaH$_2$ in the simple hexagonal structure at 60~GPa have been
calculated as being 0.22 and 780~K, respectively, giving rise to
$T_c$ on the order of only a few mK \cite{Tse:2009a}.

Given the recent interest in compressed hydrogen--rich solids as
potential superconductors, we began to wonder if H$_2$ may mix with
\mgh2\ under pressure and what other metallic systems may be found. As
will be shown in a moment, while the $\zeta$ phase is preferred for
\mgh2\ above 165 GPa, it is not the most stable point on the hydrogen--rich Mg/H phase diagram at these pressures.

\subsection{Stabilization of the Polyhydrides}

The first ionization--potential of magnesium is $\sim$2~eV larger than
that of lithium, suggesting that stabilization of the magnesium
polyhydrides will occur at a somewhat higher pressure than the lithium
polyhydrides (that is, above 100~GPa \cite{Zurek:2009c}). The ionic
radius of Mg$^{2+}$ is about 30\% smaller than that of Ca$^{2+}$, and
since CaH$_4$ was found to be the most stable stoichiometry at
$P=50-150$~GPa and CaH$_6$ at 200~GPa \cite{Ma:2012b}, we may expect
that the most favorable MgH$_n$ combination has $2 < n < 6$. How do
these predictions --- based upon chemical intuition that was developed
by analyzing computational studies of compressed polyhydrides
\cite{Zurek:2009c, Zurek:2011d, Zurek:2011h, Zurek:2012b, Zurek:2012g,
  Zurek:2012n, Zhou:2012a, Ma:2012b} --- compare with the results from our
evolutionary structure searches?
\begin{figure}[b!]
 \centering
   \includegraphics[width=\columnwidth]{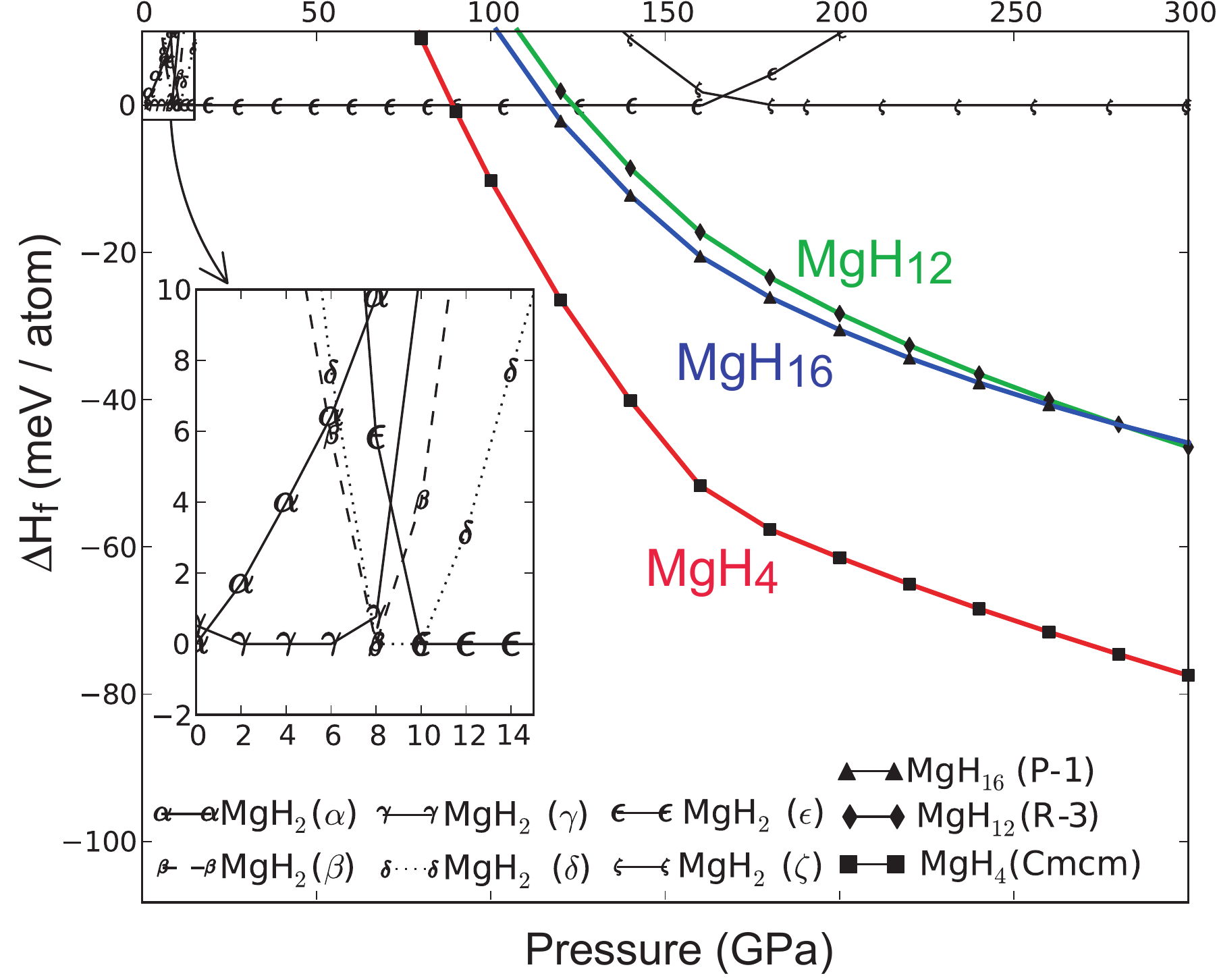}
\caption{$\Delta H_{F}$ of the MgH$_4$, MgH$_{12}$, and MgH$_{16}$
  phases as a function of pressure. The plot also illustrates the
  enthalpy of the various MgH$_2$ phases with respect to the most
  stable structure at a given pressure, with the region below 15~GPa
  magnified in the inset.} \label{fig:tieline}
\end{figure}

The calculated enthalpies of formation, $\Delta H_F$, of the important
MgH$_n$ structures found in our searches are provided in
Fig.\ \ref{fig:tieline} (see the SI for the plot of $\Delta H_F$
vs.\ H$_2$ composition). At 100~GPa only MgH$_4$ is predicted to
resist decomposition into MgH$_2$ and H$_2$. In fact, MgH$_4$ persists
as having the most negative $\Delta H_F$ as the pressure is
increased. By 200~GPa all stoichiometries except for MgH$_{2.5}$,
MgH$_3$ and MgH$_{3.5}$ have a negative $\Delta H_F$. Because MgH$_4$,
MgH$_{12}$ and MgH$_{16}$ lie on the convex hull at 200~GPa, which is
provided in the SI, they are thermodynamically stable with respect to
decomposition into other phases and
MgH$_2$/H$_2$. Fig.\ \ref{fig:tieline} illustrates that whereas the
$\Delta H_F$ of MgH$_4$ starts to become negative near 92~GPa, for
MgH$_{12}$ and MgH$_{16}$ this occurs at 122 and 117~GPa,
respectively.  Comparison of these findings with the predictions based
upon our newly developed chemical intuition under pressure illustrates
that we did reasonably well in predicting the stabilization pressures
and most stable stoichiometry without performing any
computations. Now, let us begin our exploration of the structural
peculiarities and electronic structures of the magnesium polyhydrides
falling on the convex--hull.

\subsection{MgH$_4$: H$_2$ molecules and hydridic atoms}
\begin{figure}
 \centering
   \includegraphics[width=\columnwidth]{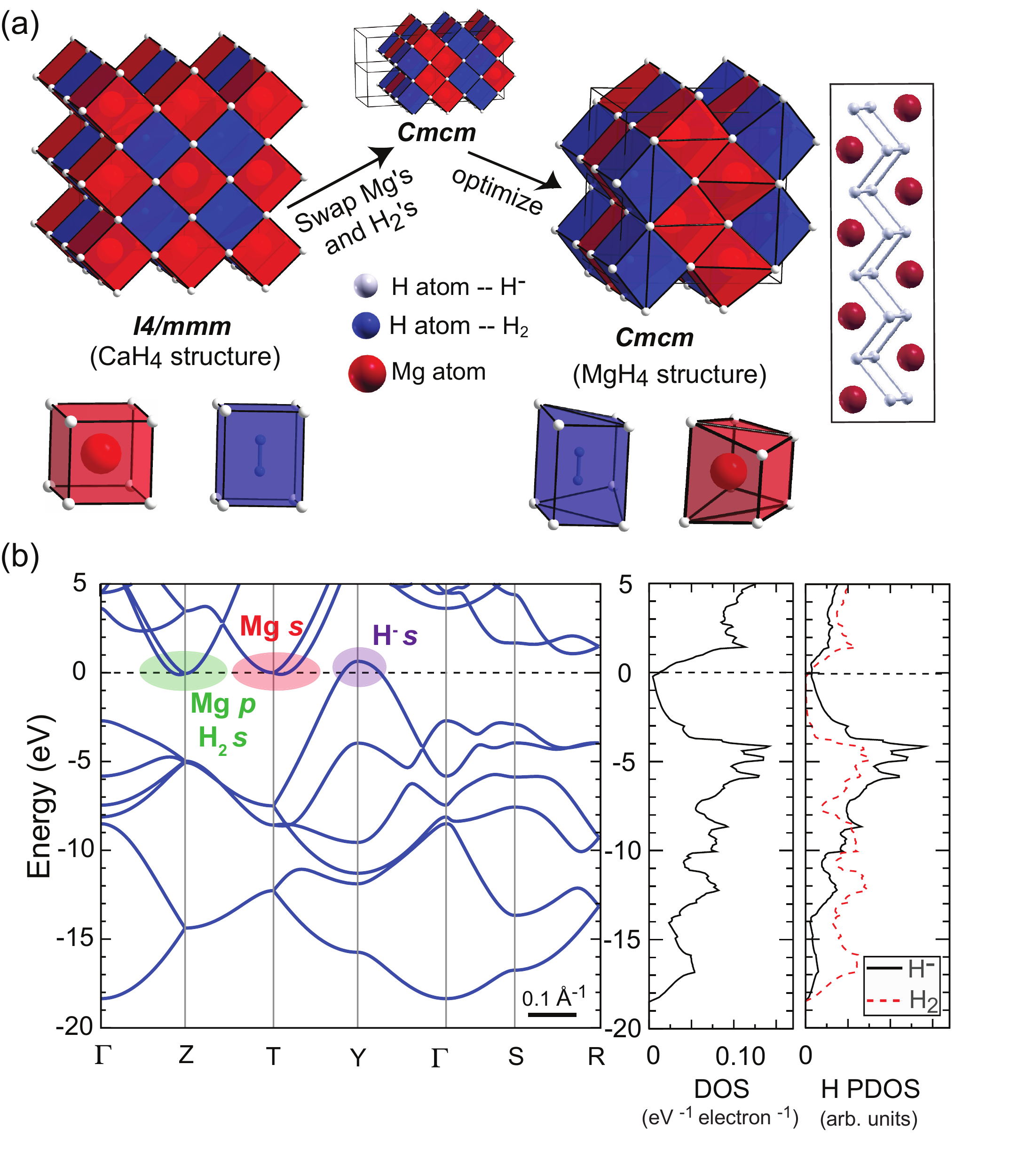}
\caption{(a) A schematic illustration of the structural similarity
  between $Cmcm$--MgH$_4$ (right) and the $I4/mmm$--symmetry CaH$_4$
  structure from Ref.\ \cite{Ma:2012b} (left). The hydride cages
  around Mg$^{2+}$/Ca$^{2+}$ and the H$_2$ molecules are colored red
  and blue, respectively \cite{endeavor}.  (b) The PBE electronic band
  structure of MgH$_4$ at 100~GPa. The character of the bands which
  cross the Fermi level is highlighted. Also shown on the right are
  the total electronic and hydrogen site--projected densities of
  states (DOS).} \label{fig:MgH4:bandStruct}
\end{figure}

The phase with the most negative enthalpy of formation throughout
the pressure range studied, $Cmcm$--\mgh4, forms a distorted CsCl bcc
lattice with H$^-$ atoms taking up the vertex positions of the
underlying cube--like cages. The body--centered positions are occupied
by a 1:1 mix of Mg$^{2+}$ cations and H$_2$ molecules with slightly
elongated bonds, as highlighted by the red and blue polyhedra in
Fig.\ \ref{fig:MgH4:bandStruct}(a). The size mismatch between the two
results in the hydrides forming different sized cages around them. The
same $Cmcm$--\mgh4 structure was discovered in evolutionary runs
carried out at 100 and 200~GPa, and phonon calculations at 100~GPa
showed it was dynamically stable.

Interestingly, the lowest enthalpy $I4/mmm$--symmetry CaH$_4$
structure found using the particle--swarm optimization technique
\cite{Wang:2010a} in a recent study \cite{Ma:2012b} can be described
the same way as $Cmcm$--\mgh4. The difference between the two is the
manner in which the cations and the H$_2$ molecules are distributed:
in CaH$_4$ they are dispersed homogeneously, but in MgH$_4$ they are
arranged into interwoven zig--zag chains. Additionally, the hydride
cages are more distorted in the MgH$_4$ structure with the smallest
face which surrounds an Mg$^{2+}$ being the one which links one
magnesium cation to another. Actually, the structure can also be
viewed as sheets of MgH$_2$ with \hydrogen\ molecules trapped in the
larger gaps between the sheets, one such sheet is shown at the right
of Fig.\ \ref{fig:MgH4:bandStruct}(a).

Within PBE the total density of states of $Cmcm$--\mgh4 at 100~GPa is
quite small; in fact it is lower than that of the $I4/mmm$ analogue. The
former is 42~meV/atom more stable than the latter at this pressure,
but the volume of $I4/mmm$--MgH$_4$ is 0.04~\AA$^3$ per atom smaller
than that of $Cmcm$--\mgh4. The formation of the zig--zag like chains
illustrated in Fig.\ \ref{fig:MgH4:bandStruct}(a) must therefore have
a substantial impact on the electronic contribution to the enthalpy.

As the pressure increases from 100 to 300~GPa, the shortest distance
between the metal cations decreases from 2.56~\AA{} to 2.32~\AA{}, and
the Mg $2p$ core bands broaden from 0.19 to 0.58~eV due to core
overlap.  The shortest Mg--\hydride\ distance decreases from
1.74~\AA{}\ to 1.55~\AA{}, as does the Mg--\hydrogen\ separation
(1.92~\AA{}/1.68~\AA{}). The molecular \hydrogen\ bonds lengthen
somewhat from 0.78~\AA{}\ to 0.79~\AA{}. This is slightly longer than
the intramolecular distance in pure compressed \hydrogen\ at these
pressures, 0.73~\AA{}\ and 0.75~\AA{}, respectively.

Since MgH$_4$ contains H$^-$ units, it becomes metallic as a result of
pressure induced band broadening and eventual overlap, like LiH$_2$
and NaH$_9$. Metallization occurs within PBE already by 20~GPa, even
though $Cmcm$--\mgh4\ is not stable with respect to decomposition into
H$_2$ and \mgh2\ at this pressure. A band displaying H$^-$
$s$--character rises above $E_F$ at the $Y$--point, whereas bands
exhibiting predominantly Mg $sp$ and a little bit of H$_2$
$\sigma^*$--character fall below it at the $Z$ and $T$--points (see
Fig.\ \ref{fig:MgH4:bandStruct}(b), and the ``fat bands'' in the
SI). Calculations using the HSE06  functional increase the metallization pressure to $\sim$150~GPa (see the SI).

\subsection{MgH$_{12}$ and MgH$_{16}$: the H$_2^{\delta -}$ belt}

The absence of H$^-$ anions in the other structures lying on the
convex hull at 200~GPa, \mgh{12}\ and \mgh{16}, has important
consequences for their electronic structure.  The hydrogenic
sublattices of these phases are composed solely of H$_2^{\delta-}$
molecules with slightly stretched bonds, just like the CaH$_{12}$
structure predicted in a previous study \cite{Ma:2012b}. The presence
of hydridic hydrogens in systems with a small mole fraction of
hydrogen, and their absence in phases with a high hydrogen content was
rationalized by Wang and co--workers by considering the formal number
of `effectively added electrons' (EAE) which are donated from the
alkaline earth metal valence bands into the H$_2$ $\sigma^*$--bands
\cite{Ma:2012b} for various CaH$_n$ stoichiometries. If the EAE is
small, the dihydrogen bond simply stretches as a result of the
population of H$_2$ $\sigma^*$, and hydridic hydrogens are not
formed. If the EAE is large enough, the molecule dissociates into
two H$^-$ units.

\begin{figure}
\centering
\includegraphics[width=\columnwidth]{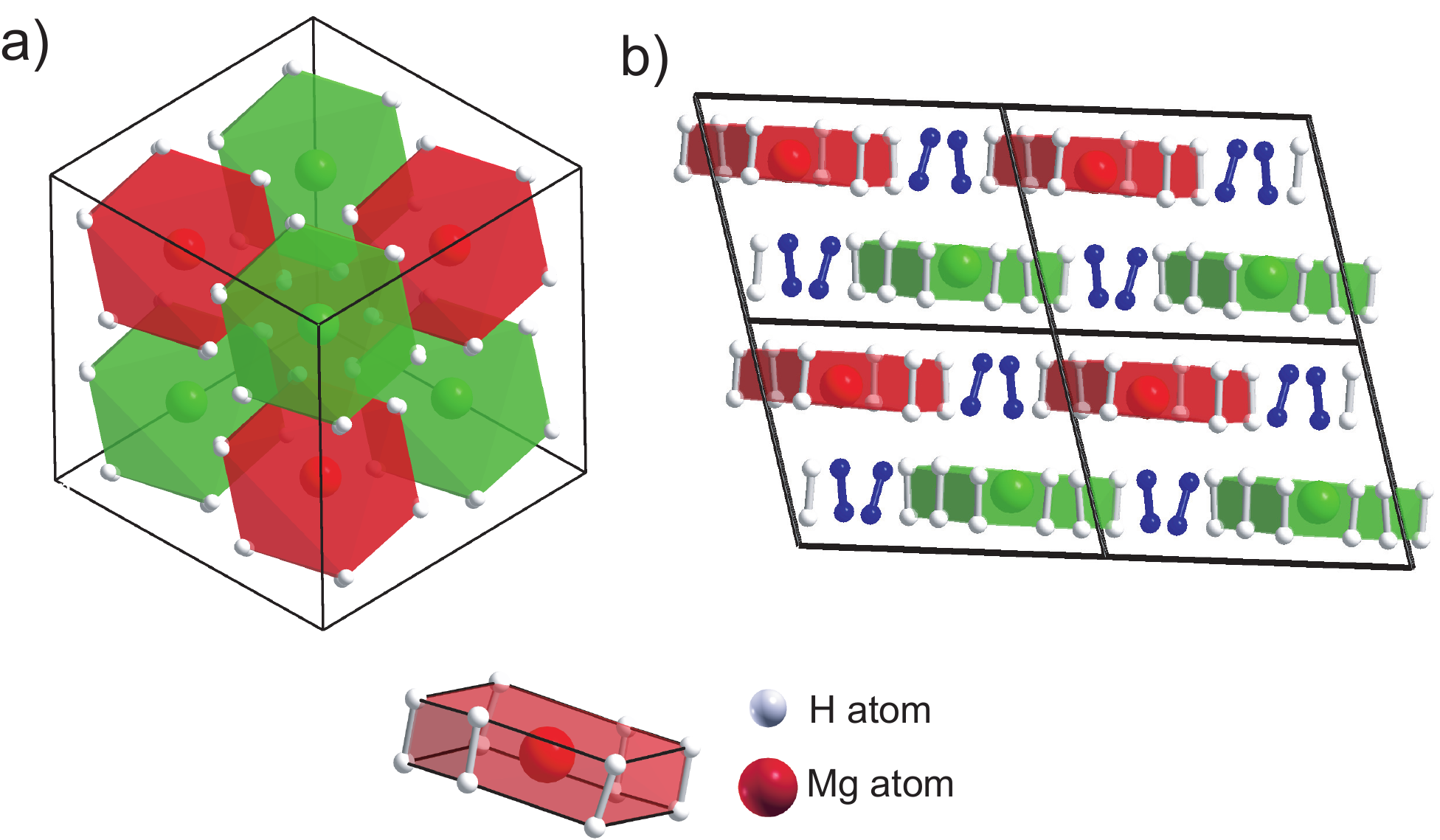} 
\caption{(a) A $2\times2\times2$ supercell of $R3$--\mgh{12}\ where
  the polyhedral units are highlighted to emphasize the
  ABCABC... close-packed arrangement. The MgH$_{12}$ building blocks
  are colored such that the red polyhedra all lie in the same
  plane. (b) A supercell of $P-1$--\mgh{16}. The \mgh{12} building
  blocks are colored to emphasize the similarity to the structure in
  (a). The blocks colored red or green lie in the same plane, and the
  hydrogens colored in blue do not belong to an \mgh{12} building
  block \cite{endeavor}.
}  \label{fig:MgH12:struct}
\end{figure}

The most stable \mgh{12} structure found in our evolutionary searches
had $R3$--symmetry and was shown to be dynamically stable at
140~GPa. The hexagonal building block for this phase consists of an
Mg$^{2+}$ cation surrounded by six H$_2^{-1/3}$ molecules, as
illustrated in Fig.\ \ref{fig:MgH12:struct}(a). The dihydrogen
molecules in these hexagons appear to arrange in a `belt' around the
metal cation in a side--on fashion. The hexagons are tiled in parallel
sheets forming an ABCABC... close--packed structure. This description
is, however, somewhat misleading as the distance between Mg$^{2+}$ and
a hydrogen atom within the \mgh{12}\ building block is comparable to
the shortest distance between the metal ion and a hydrogen atom in a
different building block. For example, the intra-- and inter--building
block Mg--H measurements are 1.76 and 1.81 \AA{}, respectively, at
140~GPa.  The dihydrogen bonds in \mgh{12} are substantially elongated
to 0.82 and 0.81~\AA{} at 140 and 300~GPa, when compared with the
typical bond lengths of pure \hydrogen\ at these pressures (0.74/0.75
\AA{}). A gas phase geometry optimization on a H$_2^{-1/3}$ molecule
at 1~atm showed that the bond stretched from 0.75 to 0.79~\AA{}. So
the expansion of the H--H bond observed in \mgh{12} is consistent with
the partial filling of the \hydrogen\ $\sigma^*$--bands.

The lowest enthalpy \mgh{16} configuration we found, which exhibited
$P-1$ symmetry and was dynamically stable at 130~GPa, could be thought
of as being made up of similar \mgh{12} units with excess
H$_2^{\delta^-}$ molecules stuffed between the hexagons, as
illustrated in Fig.\ \ref{fig:MgH12:struct}(b). These `interstitial'
dihydrogens lie on top of the Mg$^{2+}$ cations in the \mgh{12}
building blocks. The intramolecular distance in H$_2^{-1/4}$ at 1~atm
was calculated as being 0.78~\AA{}, whereas the bond lengths of the
dihydrogens in \mgh{16} range from 0.76 to 0.85~\AA{} at 130~GPa, with
the interstitial hydrogen molecules being shorter (0.76-0.78~\AA{})
than those in the belt (0.79-0.85~\AA{}). The unequal bond lengths in
the H$_2^{\delta^-}$ units may be a result of different charge states,
or be due to their local bonding environments, or both
\cite{Labet:2012a}.  As the pressure increases, the intramolecular
H--H bond lengths decrease, with the interstitial and belt hydrogens
measuring 0.74-0.76~\AA{} and 0.78-0.84~\AA{}, respectively, at
300~GPa.

\begin{figure}
 \centering
   \includegraphics[width=\columnwidth]{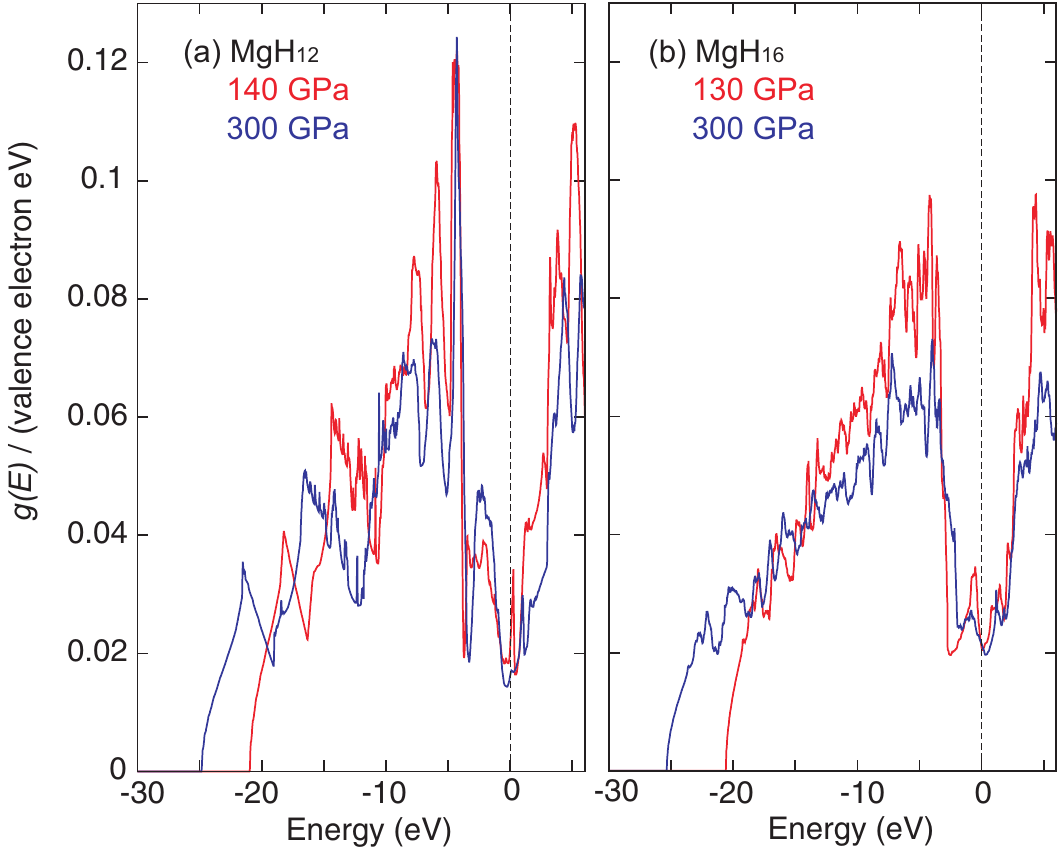} 
  \caption{The PBE valence densities of states of (a) MgH$_{12}$, and
    (b) MgH$_{16}$. Note that the Fermi level lies just below
    (directly after) a sharp peak in the DOS in MgH$_{12}$ at 140~GPa
    (MgH$_{16}$ at 130~GPa).  } \label{fig:dos}
\end{figure}

Since \mgh{12} and \mgh{16} are metallic as a result of the partial
filling of the H$_2$ $\sigma^*$--bands like LiH$_6$, they have a high
density of states at $E_F$, see Fig.\ \ref{fig:dos}. Both phases
remain good metals upon compression up to at least 300~GPa, and the Mg
$2p$ bands broaden only slightly to $\sim$0.15-0.2~eV at 300~GPa as a
result of core overlap. A comparison of the DOS of \mgh{12} at 140~GPa
calculated with PBE and the HSE06 screened hybrid functional showed
that latter valence DOS was slightly broader (as expected for metallic systems \cite{Biller:2011a, Paier:2006a}), and the core Mg
$2p$ bands shifted to lower energies. However, the $g(E_F)$ computed
with the two functionals was essentially the same (see the SI). This
is in--line with our previous results which showed that $g(E_F)$ is
relatively insensitive to the choice of the functional in sodium
polyhydrides that did not contain hydridic hydrogens
\cite{Zurek:2011d}. The nearly free--electron like valence bands of
\mgh{12} and \mgh{16} exhibit primarily H $s$--character, with a
little bit of Mg $sp$ spanning throughout (see the fat bands in the
SI). Nonetheless, the computed DOS of these structures match quite
well with a hypothetical system where the metal cations have been
removed, the (H$_{12}$)$^{2-}$ and (H$_{16}$)$^{2-}$ lattices (see the
SI), suggestive of almost full ionization of the Mg valence $3s$
electrons into the H$_2$ $\sigma^*$--bands.

\subsection{The MgH$_{12}$ Building Blocks}
A number of MH$_{12}$ clusters, where M is a transition metal atom,
have been predicted as being stable in the gas phase by quantum
chemical calculations \cite{Gagliardi:2004a}. Moreover
WH$_4$(H$_2$)$_4$, which contains four hydridic hydrogens and four
dihydrogens bonded to the metal center in a side--on fashion, has been
made in a neon matrix \cite{Wang:2008a}. Unsurprisingly, molecular
calculations on both the optimized MgH$_{12}$ building blocks as well
as (MgH$_{12}$)$^{2+}$ with $D_{6h}$ symmetry revealed numerous
imaginary frequencies, so these clusters will not be stable at
1~atm. Nonetheless, it is instructive to compare the electronic
structure of the MgH$_{12}$ building block with the DOS computed for
$R3$--MgH$_{12}$ at various pressures.
\begin{figure}
\centering
\includegraphics[width=1.0\columnwidth]{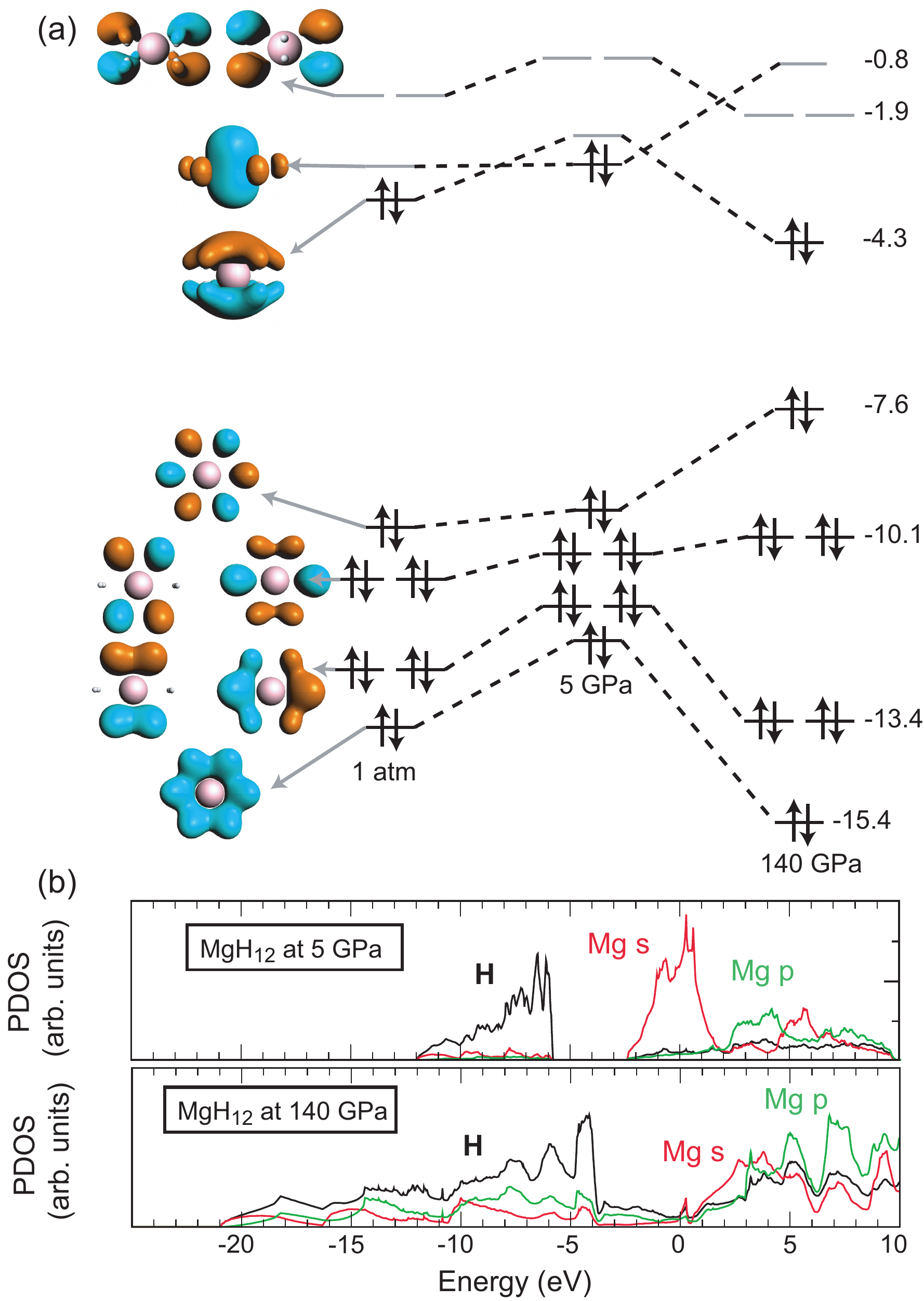}
\caption{(a) Calculated molecular orbital level diagram of an
  MgH$_{12}$ cluster when optimized in the gas phase at 1~atm (left)
  and in the geometry which it adopts in the $R3$--MgH$_{12}$ solid at
  5~GPa (center) and 140~GPa (right). The energies of the MOs at
  140~GPa are given to the right in eV. (b) Calculated site--projected
  densities of states of $R3$--MgH$_{12}$ at 5 and
  140~GPa.}  \label{fig:cluster}
\end{figure}

Whereas in an optimized MgH$_{12}$ cluster the Mg--H and
intramolecular H--H distances were calculated as being 2.20 and
0.79~\AA{}, these values measure 2.34-2.58/1.76-1.78~\AA{} and
0.79/0.82~\AA{} in $R3$--MgH$_{12}$ at 5/140~GPa. So the gas phase
cluster is actually a little bit more `compressed' than the building
block within the solid at 5~GPa. The energy level diagrams and
canonical molecular orbitals (MOs) of these clusters are illustrated
in Fig.\ \ref{fig:cluster}(a). The six MOs lowest in energy resemble
the canonical MOs of benzene, except that they do not contain a node
which bisects the dihydrogen molecules. A large gap is found between
the highest occupied molecular orbital (HOMO) and HOMO-1 of the
optimized cluster, but a small gap separates the HOMO and lowest
unoccupied molecular orbital (LUMO). The HOMO--HOMO-1 gap decreases as
the cluster is compressed, as do the analogous sets of bands in the
extended system, as illustrated in Fig.\ \ref{fig:cluster}(b).

The frontier orbitals of the 1~atm cluster contain substantial
character arising from H, and the HOMO also has an important
contribution from the Mg $3p$ and the LUMO from the Mg $3s$
orbitals. Our fragment orbital analysis \cite{Velde:2001a} shows that
the LUMO+1 contains primarily H--character with less than 4\% of Mg
$3d$ states mixed in. In the `less--compressed' cluster extracted from
the solid at 5~GPa the frontier orbitals swap positions so that the
HOMO displays Mg $3s$ and the LUMO Mg $3p$--character. This is
in--line with the projected densities of states which illustrate that
the predominant contribution around the Fermi level is due to Mg $s$
at 5~GPa. At 140~GPa the gap between the hydrogenic and the metallic
bands between -6 to -3~eV closes in the solid, and the HOMO and HOMO-1
orbitals in the cluster come closer together in energy. Moreover, as a
result of the pressure induced $s\rightarrow p$ transition in Mg, the
states around the Fermi level contain about an equal amount of Mg $s$
and $p$--character. In the cluster the energy ordering of the
molecular orbitals also changes, so that the HOMO is more Mg
$p$--like, and the LUMO displays primarily H $s$ character. So the gas
phase clusters are able to mimic some of the essential features of the
projected DOS of the solid at different pressures.

\subsection{Superconductivity in the polyhydrides}

Our computations have estimated the $T_c$ of the classic alkaline
earth hydride, $P6_3/mmc$--MgH$_2$, as being $\sim$20~K at 180~GPa. We
wondered what the $T_c$ of the aforementioned hydrogen--rich phases
would be in comparison, and if $T_c$ would be influenced by the
hydrogenic sublattice of the polyhydride. Computations have been
carried out on $Cmcm$--MgH$_4$ and $R3$--MgH$_{12}$ at 100 and
140~GPa, respectively, since these are pressures slightly larger than
when $\Delta H_F$ was computed as becoming negative, and since both
systems were found to be metallic within PBE at these pressures.

\begin{figure}
\centering
\includegraphics[width=\columnwidth]{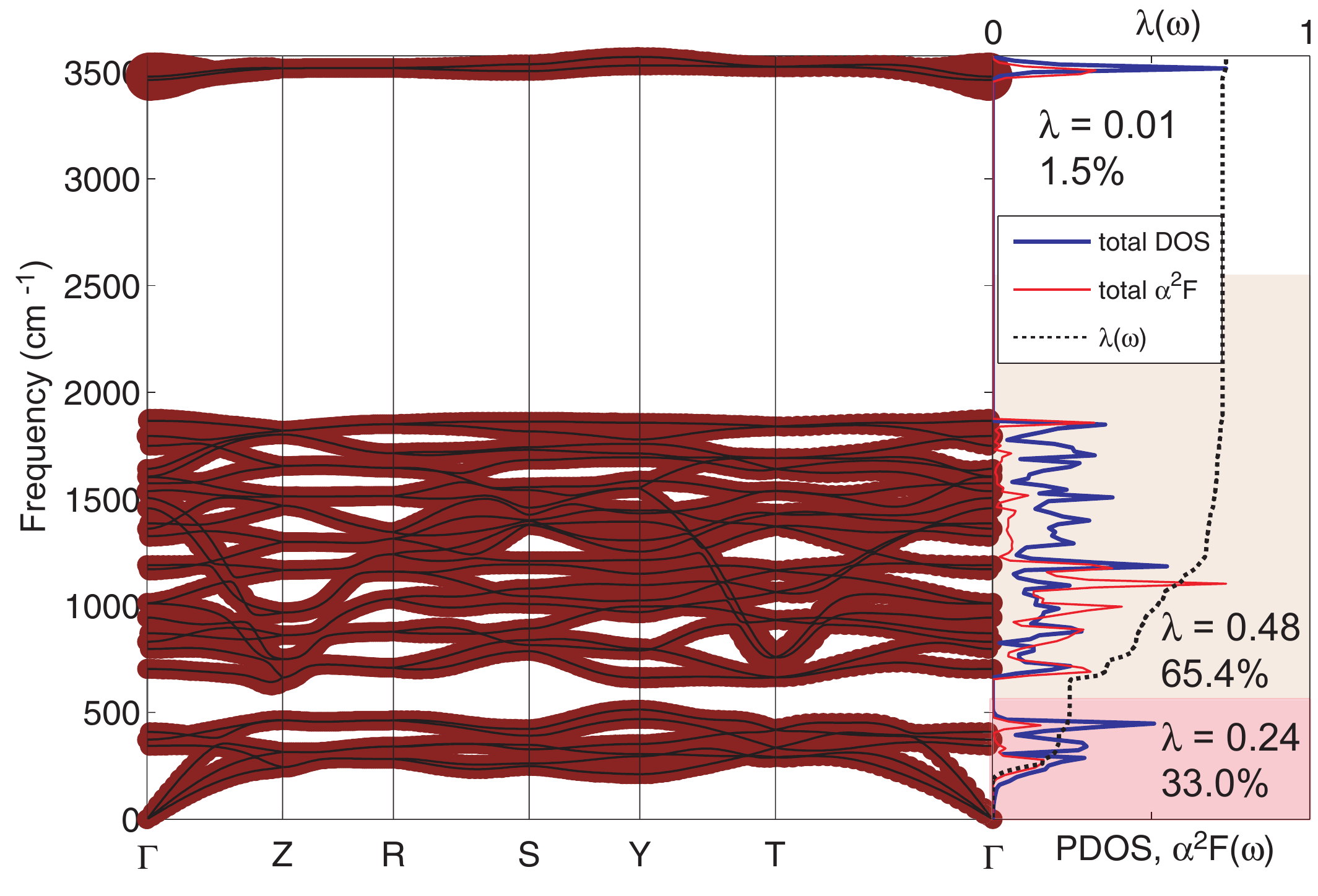}
\caption{Phonon band structure, phonon density of states and the
  Eliashberg spectral function, $\alpha^{2}$F($\omega$), of MgH$_{4}$
  at 100~GPa. Circles indicate the phonon linewidth with a radius
  proportional to the strength.  At this pressure $\lambda=0.74$,
  $\omega_{log}=941$~K, and $T_c=29-37$~K assuming
  $\mu^*=0.13-0.1$.} \label{fig:MgH4:critical}
\end{figure}

Within PBE the density of states at the Fermi level of $Cmcm$--MgH$_4$
at 100~GPa is comparable to the one calculated for \zmgh\ at
180~GPa. The total EPC parameter is  nearly 28\%
larger, however. In both phases the low--frequency modes which are
mostly due to the vibrations of the Mg atoms contribute $\sim$0.2 to
the total $\lambda$; compare Fig.\ \ref{fig:MgH4:critical} and
Fig.\ \ref{fig:MgH2:critical}. The main reason why the overall EPC parameter is
larger for the polyhydride than the classic hydride is a result of the
total coupling provided by the modes between $\sim$500-2500~cm$^{-1}$,
which are primarily due to the motions of the hydrogen atoms. Whereas
the classic hydride contains only hydridic hydrogens, there is an
equal number of H$_2$ and H$^-$ hydrogens in MgH$_4$. The H$_2$
vibron, located around 3500~cm$^{-1}$ contributes only 1.5\%
to the total coupling strength in the polyhydride. This is not
surprising, since the bands crossing the Fermi level displayed only a
small amount of H$_2$ $s$--character. Despite the higher $\lambda$
that MgH$_4$ has as compared to MgH$_2$, the prefactor in the modified
McMillan equation, $\omega_\text{log}$, is calculated as being 15\%
smaller so the total $T_c$ of MgH$_4$ is estimated as being
$\sim$14~K higher than that of the classic alkaline earth hydride.

\begin{figure}[h!]
 \centering
 \includegraphics[width=\columnwidth]{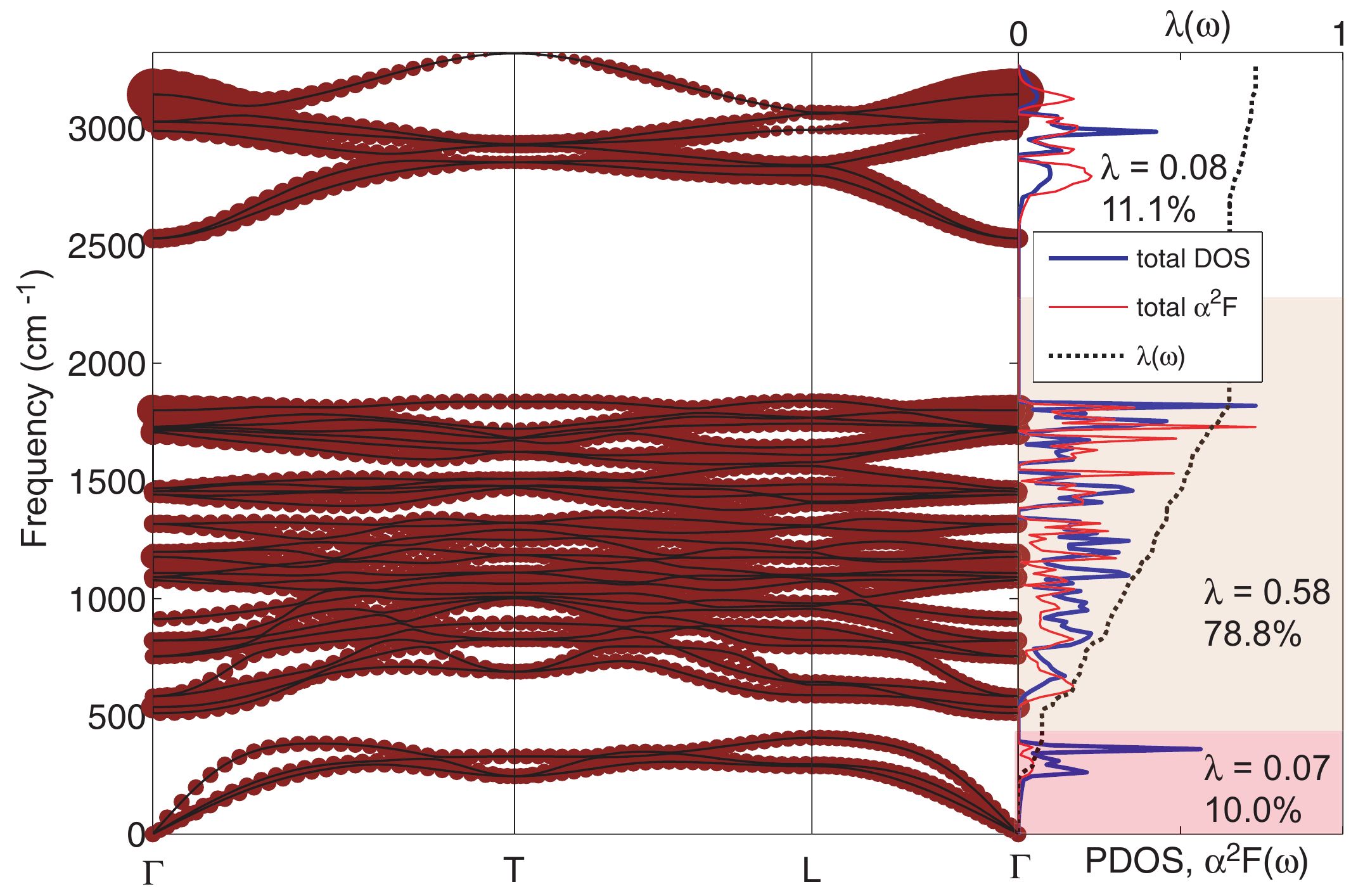}
\caption{Phonon band structure, phonon density of states and the
  Eliashberg spectral function, $\alpha^{2}$F($\omega$), of MgH$_{12}$
  at 140~GPa. Circles indicate the phonon linewidth with a radius
  proportional to the strength.  At this pressure $\lambda=0.73$,
  $\omega_{log}=1554$~K, and $T_c=47-60$~K assuming $\mu^*=0.13-0.1$.}
\label{fig:MgH12:critical}
\end{figure}

Despite the increased hydrogen content and higher density of states at
the Fermi level, the total EPC parameter of $R3$--MgH$_{12}$ at 140~GPa is
almost the same as of $Cmcm$--MgH$_4$ at 100~GPa. However, a
comparison of the $\lambda(\omega)$ in Fig.\ \ref{fig:MgH12:critical}
and Fig.\ \ref{fig:MgH4:critical} shows that the relative
contributions to the overall $\lambda$ are quite different in the two
polyhydrides.  MgH$_{12}$ does not contain any hydridic hydrogens, and
is metallic because of the partial filling of the H$_2$
$\sigma^*$--bands. In contrast to what was found for MgH$_4$, the
H$_2$ vibron contributes about 11.1\% to the total $\lambda$.  This
corresponds quite well to the 11.3\% calculated for a compressed KH$_6$ phase \cite{Zhou:2012a} whose hydrogenic sublattice only contained
H$_2^{\delta -}$ motifs. The EPC associated with the low--frequency
modes below 400~cm$^{-1}$, which are dominated by the motions of the
heavier metal atoms, is about a third of the amount calculated for
the phases containing H$^-$ units. The main contribution to $\lambda$,
79\%, arises from the intermediate frequency regime, which is
primarily due to the H$_2$ motions. The reason why the $T_c$ of
MgH$_{12}$ is estimated as being $\sim$20~K higher than that of
MgH$_4$ is due to the larger $\omega_\text{log}$. Unfortunately, the computational
expense precluded us from calculating the EPC parameter of compressed
$P-1$--MgH$_{16}$, or from exploring the pressure dependence of $T_c$.

The values we calculate for the total EPC parameter and the $T_c$ of
the magnesium polyhydrides falls within the range of 0.5-1.6 and
10-139~K, respectively, computed for a number of hydrogen--rich systems
\cite{Ma:2012b}. We show that comparable $\lambda$ values may be obtained for polyhydrides with very
different hydrogenic sublattices, but their $\omega_\text{log}$ and therefore $T_c$ may differ. The magnesium polyhydrides are predicted to have a larger $T_c$ than MgH$_2$ under pressure, and phases with a larger mole percent ratio will likely have a higher 
$T_c$.

\section{Conclusions}

Evolutionary structure searches coupled with density functional theory
calculations are used to predict the most stable structures and
stoichiometries of the magnesium polyhydrides, MgH$_n$ with $n \ge 2$,
under pressure. The thermodynamically stable structures found in this study have a hydrogenic sublattice containing H$^-$ anions and H$_2$ units (MgH$_4$), or H$_2$ molecules which are less strongly bonded than those found in pure molecular hydrogen at 1~atm (MgH$_{12}$ and MgH$_{16}$).

Metallization in $P6_3/mmc$--MgH$_2$ occurs as a
result of an H$^-$ $s$--band rising above, and a Mg $sp$ band falling
below the Fermi level. $T_c$ is estimated as being between 16-23~K at
180~GPa, with a sizable contribution to the total electron phonon coupling parameter arising from vibrations related to both the hydrogen and magnesium atoms. 

MgH$_4$, which starts to become thermodynamically stable with respect
to decomposition into MgH$_2$ and H$_2$ near 100~GPa is found to
contain one H$_2$ molecule and two hydridic hydrogens per Mg$^{2+}$
cation. Metallization occurs as a result of pressure--induced band gap
closure, but the density of states at the Fermi level is quite
low. Around 120~GPa other stoichiometries, whose hydrogenic
sublattices contain only H$_2^{\delta-}$ molecules with slightly
stretched bonds, emerge as being thermodynamically stable. MgH$_{12}$
and MgH$_{16}$ are metallic in part as a result of the partial filling
of the H$_2$ $\sigma^*$--bands and have a high density of states at
the Fermi level. Their electronic structure at various pressures can
be traced back to the molecular orbital diagram of their building
block, the MgH$_{12}$ cluster. Despite the very different hydrogenic sublattices, both MgH$_4$ and MgH$_{12}$ are found to have similar electron phonon coupling parameters. The main reason why the $T_c$ of MgH$_{12}$ at 140~GPa is calculated as being larger than that of MgH$_4$ at 100~GPa, 47-60~K vs.\ 29-37~K assuming typical values of $\mu^*$, is because of the larger average logarithmic frequency computed for MgH$_{12}$.

\begin{acknowledgements}
We acknowledge the NSF (DMR-1005413) for financial support, and the
Center for Computational Research at SUNY Buffalo for computational
support. B.A.\ was supported by a postdoctorate scholarship of CoHE
(Turkish Council of Higher Education), and acknowledges
ULAKBIM-TR-Grid for computational time.
\end{acknowledgements}


\end{document}